\documentclass[preprint,10pt,aps,twocolumn,amsmath,amssymb, nofootinbib,superscriptaddress,hyperref,showkeys]{revtex4}
\usepackage[utf8]{inputenc}

\usepackage{array} 
\usepackage{amssymb,amsmath} 
\usepackage{tikz}
\usepackage{hyperref}  
\usepackage{slashed} 

\newcommand{\beq}{\begin{equation}}
\newcommand{\eeq}{\end{equation}}
\newcommand{\bea}{\begin{eqnarray}}
\newcommand{\eea}{\end{eqnarray}}
\newcommand{\beas}{\begin{eqnarray*}}
\newcommand{\eeas}{\end{eqnarray*}}
\newcommand{\bi}{\begin{itemize}}
\newcommand{\ei}{\end{itemize}}

\def\tev{\,{\ifmmode\mathrm {TeV}\else TeV\fi}}
\def\gev{\,{\ifmmode\mathrm {GeV}\else GeV\fi}}
\def\to{\rightarrow}

\def \pslash{\rm p{\!\!\!/}}
\begin{document}
	
\author{Akanksha Bhardwaj}
\email{akanksha@prl.res.in}
\affiliation{Theoretical Physics Division, Physical Research Laboratory, Ahmedabad 380 009, India}

\author{Partha Konar}
\email{konar@prl.res.in}
\affiliation{Theoretical Physics Division, Physical Research Laboratory, Ahmedabad 380 009, India}

\author{Pankaj Sharma}
\email{pankaj.sharma@adelaide.edu.au}
\affiliation{Center of Excellence for Particle Physics at the Terascale, University of Adelaide,  SA 5005, Australia}

\author{Abhaya Kumar Swain}
\email{abhaya@prl.res.in}
\affiliation{Theoretical Physics Division, Physical Research Laboratory, Ahmedabad 380 009, India}
\affiliation{School of Physical Sciences, Indian Association for the Cultivation of Science, Kolkata 700 032, India}

\title{Exploring CP phase in $\tau$-lepton Yukawa coupling in Higgs decays at the LHC}

%
%
%

\begin{abstract}
We study the prospect of determining the CP violating phase in $\tau$-lepton Yukawa coupling at the Large Hadron Collider (LHC). While the current run is already exploring the production of a pair of the third generation $\tau$ leptons from Higgs decay, these measurements are not sensitive enough to constrain the CP violating phase. In this paper, several CP odd observables are proposed and analyzed utilizing the dominant channels with the semi-invisible hadronic decay of $\tau$. Several asymmetries corresponding to the T odd momentum correlations are also studied and their sensitivities to the CP violating phase in tau-lepton Yukawa couplings are estimated at 13 TeV LHC with 1000 ${fb}^{-1}$ of integrated luminosity. We also present a novel way to reconstruct $\tau$ momentum at the LHC utilizing the information of impact parameter. Finally, we obtain that the asymmetries can be as large as 35$\%$ for a case of maximal CP violation in the $\tau$ Yukawa couplings.
\end{abstract}


\maketitle

\section{Introduction}
\label{sec:intro}
The Large Hadron Collider (LHC) has achieved a milestone when it discovered a Standard Model (SM) like boson with mass around 125 GeV in its runs at 7 and 8 TeV~\cite{Chatrchyan:2012ufa,:2012gk}. With the current data, the accuracy is still not adequate enough to confirm/refute it to be the SM Higgs. The data as of now allows for the significant deviations in Higgs couplings with the fermions and gauge bosons from the SM predictions. Thus there remains sufficient scope for new physics in the Higgs sector. In the current and future runs, precise determination of Higgs couplings with the SM fermions is one of the foremost goals for the LHC. In this context, owing to their large Yukawa couplings, the Higgs couplings with the third generation fermions {\it viz.} $Ht\bar t$ and $H\tau^+\tau^-$ become crucial. Moreover, heavier fermions from the third generation held the clue to the electroweak symmetry breaking (EWSB), and thus are expected to shed light on different aspects  such as, coupling structure and CP properties of the resonant state. In a particular scenario with beyond the SM (BSM), new physics effects contributing in these couplings can also be  constrained further.

Another intriguing open question in particle physics is the CP violation (CPV) which still lacks a full understanding. CPV has been experimentally found and extensively studied in the mixings and decays of $K$ and $B$ systems \cite{Christenson:1964fg,Abe:2001xe}. It also plays a crucial role in observed baryon asymmetry of the universe. In fact, it is one of the three Sakharov's conditions to explain the asymmetry. In the SM, the only source of CPV is the phase associated with the Cabibbo-Kobayashi-Maskawa (CKM) inter-generational quark mixing matrix. However, the amount of CPV present in the SM cannot adequately explain the baryon asymmetry. The Higgs sector of the SM is CP conserving as all the Higgs couplings are CP even. However, extensions of the SM like two-Higgs doublet models (2HDM) and minimal supersymmetric standard model (MSSM) contain two Higgs doublets which lead to additional bosons in the model. One of these bosons lead to CP odd couplings with the SM particles. In the CP violating scenarios of these models, all the three scalars mix with each other leading to the CP violating couplings with other particles in the model~\cite{Pilaftsis:1998dd,Pilaftsis:1999qt}.

Currently the studies of the spin and parity of the Higgs boson based on the combination of channels producing EW gauge bosons $\gamma, W,Z$ point to a spin-0 particle with a pure pseudoscalar boson being ruled out at 95 \% CL~\cite{Aad:2013xqa}. However, the possibility of a CP admixture with both the scalar and the pseudoscalar components is still allowed. Thus, it would be one of the important goals of the next run of the LHC, which will be a high energy and high luminosity run, to determine the CP composition of the Higgs. A CP admixture Higgs would lead to the CP violating couplings with other SM particles. CPV in the Higgs sector is more prominent in its fermionic couplings than gauge boson couplings as the couplings of the pseudoscalar to gauge bosons are absent at tree level and can only arise at the one-loop level. Hence, the Yukawa couplings are known to be more democratic to CP even and CP odd bosons. Also, since Yukawa couplings are larger for third generation fermions and recognizing that it is difficult to study CPV in $Ht\bar t$ couplings at the LHC, in this letter we focus on the $H\tau^+\tau^-$ couplings when Higgs decays to resonant tau pairs once it is produced at the LHC. 

In literature, several observables have been proposed to measure the CPV nature in the $\tau$ Yukawa couplings at the LHC~\cite{Berge:2008wi,Berge:2008dr,Berge:2011ij,Berge:2014sra,Harnik:2013aja,Dolan:2014upa,Askew:2015mda,Hayreter:2016kyv,Han:2016bvf}. Most of these observables rely on $\tau^\pm$ polarization and require the reconstruction of full $\tau^\pm$ spin four-vector in some rest frames. In the present study, we define our observables based on definite CP and T transformation properties. These observables are momentum correlations that are constructed from visible particle momenta in $\tau^\pm$ decays. In our study, we focus on the $\tau^\pm\to \pi^\pm \nu$ decay channel which is considered to be the best channel for parity determination and has the largest polarimetric power. One can safely accommodate other dominant hadronic channel $\tau^\pm\to \pi^\pm\pi^0\nu$ using the same prescription however with reduced polarimetric power.

CMS (ATLAS) collaboration at CERN recently studied~\cite{Chatrchyan:2014nva,Aad:2015vsa} the tau pair production through Higgs decay, analyzed at center-of-mass energy 7 and 8 TeV run with corresponding integrated luminosity of 4.9 (4.5) and 19.7 (20.3) ${fb}^{-1}$ respectively. To explore and detect these $\tau$ leptons, both hadronic and leptonic decay channels are considered, resulting into multiple different final state combinations from the pair. Both of these studies reported an excess of such events over the expected backgrounds, with a local significance 3.2 (4.5) standard deviation for the 125 GeV Higgs. The measured signal strengths in both cases are consistent with the SM expectation. However, one readily identifies the relatively smaller significance in the present scenario compared to the other decay modes of the Higgs owing to the challenging final states expected from the $\tau$ pair at the LHC. In presence of multiple invisible neutrinos at the final state, reconstruction 
of these events are rather complex. 

We organize rest of the presentation as follows. In section~\ref{sec:higgs}, we introduce the importance of looking $H\to \tau^+\tau^-$ channel for the CP phase along with the corresponding Lagrangian parameterizing the CP transformation properties. Thereafter, in section~\ref{sec:CP} we introduce the CP observables and their construction which can be interesting. Since the rest frame observables are proved to be more effective, reconstruction of these semi-invisible events are necessary which will be discussed in section~\ref{sec:reconstruction}. We will also discuss a new method of reconstruction which proved to have a good efficiency for reconstructing the semi-invisible tau decay, after mentioning some existing methods.
All the results together with the capability of studying  CP phase is finally presented in section~\ref{sec:sensitivity} before  concluding in section~\ref{sec:conclusion}.

\section{Higgs boson production and decays}
\label{sec:higgs}

\begin{figure}[!t]
\includegraphics[scale=0.52,angle=0,keepaspectratio=true]{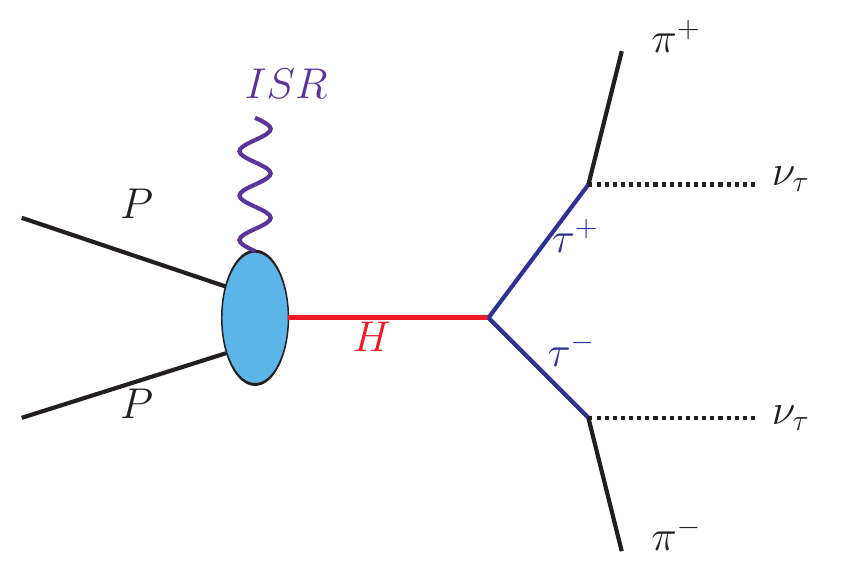}
 \caption{Representative diagram for  $h \rightarrow \tau^- \tau^+$ with tau lepton decay hadronically via one prong decay channel. We assign momenta for the final state invisible (neutrinos) and visible (pions) particles as $q_i$ and $p_i$ respectively with i = 1,2.}
 \label{fig:h2tautaudiag}
\end{figure}

In this work, we study Higgs production via gluon fusion mechanism though our methodology to determine the CP phase in the $H \tau^+ \tau^-$ coupling which can easily be applied to any other Higgs production processes, such as, VBF process or the associated vector boson productions. Following the Higgs production, we study the decay mode $H\to \tau^+\tau^-$. We consider both the $\tau^+$ and $\tau^-$ to decay hadronically in order to minimize the loss of kinematic information due to multiple missing neutrinos. For the $\tau$'s decay modes, we take into account the following 1-prong decays in our analysis: $\tau^\pm\to \pi^\pm \ \nu_\tau$. A representative diagram for the Higgs production and its decay to $\tau^+\tau^-$ followed by $\tau$ decays has been shown in Fig.~\ref{fig:h2tautaudiag}.

In our analysis, we consider the Higgs boson to be a CP admixture and does not have a definite CP transformation properties. As mentioned earlier, the model for such a scenario could be an extension of a Higgs sector such as 2HDM, MSSM etc. with a CP violation in Higgs couplings. The Yukawa terms in the Lagrangian for such a Higgs boson can be parameterized as following:
\begin{equation}\label{lag}
 \mathcal L \supset -m_\tau \bar \tau \tau - \frac{y_\tau}{\sqrt{2}}H\bar \tau (\cos\alpha+i\gamma_5\sin\alpha)\tau
\end{equation}
where $\tau$ and $H$ are the physical fields, respectively, $y_\tau$ is the effective strength of the $\tau$-Yukawa interaction and $\alpha$ denotes the degree of mixing of the scalar and pseudoscalar component of the Higgs boson. For the SM Higgs boson, $\alpha$ vanishes identically at tree level reproducing a CP even Higgs and $y_\tau=m_\tau/v$. The CP phase can vary in the range $(-\frac{\pi}{2}, \frac{\pi}{2})$ with $\alpha=\pi/2$ corresponds to a pure pseudoscalar and $\alpha=\pi/4$ to a maximally CP-violating case. Here to go forward we keep the $y_\tau$ fixed to the SM value and only vary the CP phase of the $\tau$-Yukawa coupling to study the deviations in the expectation values of the observable with respect to the CP phase. 

In our analysis, we consider Higgs mass $M_h=125$ GeV. We have incorporated the anomalous Higgs couplings to tau leptons in {\tt Madgraph}~\cite{Alwall:2014hca} using {\tt FeynRules}~\cite{Alloul:2013bka}. The decays of the taus are handled with the tau-decay model\cite{Hagiwara:2012vz} implemented in {\tt Madgraph}. We use {\tt Madgraph} to generate the parton level events which are then passed to the {\tt Pythia}~\cite{Sjostrand:2007gs} for our analysis.

\section{Observables}
\label{sec:CP}

The Higgs spin and parity information are coded into the correlations between $\tau^+$ and $\tau^-$ spins. The spin of $\tau^\pm$ and correlation between $\tau^+$ and $\tau^-$ spins are not directly measurable rather they are determined from the distribution of their decay products. They may also manifest themselves in the correlations among momenta of the $\tau^\pm$ decay products in particular to the plane transverse to $\tau^+\tau^-$ axes. This is because the decay distribution of $(H/A\to\tau^+\tau^-)$ is proportional to $d\Gamma\propto (1+s_{||}^{\tau^+}s_{||}^{\tau^-}\pm s_{\perp}^{\tau^+}s_{\perp}^{\tau^+})$~\cite{Kramer:1993jn} where $||$ and $\perp$ denote the longitudinal and transverse components of $\tau^\pm$ spin with respect to Higgs momentum as seen from the $\tau^+\tau^-$ rest frame. 

Taking into account of the aforementioned fact and recognizing that a triple product correlation is sensitive to a scalar and pseudoscalar contribution, we study several simple triple product correlations constructed out of momenta of the particles involved in the process. We utilize the momenta of the $\tau^+$, $\tau^-$ and their decay products, i.e., $\pi^\pm$, to construct momentum correlations. Under CP, $\vec{p}_{\tau^-}\xrightarrow{CP} -\vec{p}_{\tau^+}$ and $\vec{p}_{\pi^-}\xrightarrow{CP} -\vec{p}_{\pi^+}$. A triple product correlation transforms under CP as: $\vec{p}_{\tau^{-}}\cdot (\vec{p}_{\pi^{-}} \times \vec{p}_{\pi^{+}})\xrightarrow{CP} -\vec{p}_{\tau^{-}}.(\vec{p}_{\pi^{-}} \times \vec{p}_{\pi^{+}})$. Thus, all the observables listed in Table \ref{tab:obs} are CP odd and T odd\footnote{Henceforth, T will always refer to naive time reversal, i.e., reversal of all momenta and spins without interchanging the initial and final states.}. Note that the list does not exhaust all possible combination of triple product correlations involving particle momenta involved in the process.
But here our primary interest was to construct triple product in terms momenta of tau leptons and pions in the most trivial way possible. In principle, one could also include each neutrino momenta in constructing these correlations provided that they are determined at the LHC. Here we focus only on those combinations having substantial sensitivity to the cp phase.

The amplitude for the full Higgs decay chain $h\to \tau^+\tau^-\to \pi^+\pi^-\nu_\tau \bar\nu_\tau$ can be written as 
\begin{eqnarray}
 \mathcal M &\propto& \bar u_{\nu_\tau}(\pslash_{\tau^-}+m_\tau)(\cos\alpha+i\gamma_5\sin\alpha)\nonumber\\
 &\times& (-\pslash_{\tau^+}+m_\tau)P_L v_{\bar\nu_\tau}.
\end{eqnarray}
In a full matrix element squared, one would get CP angle $\alpha$ independent and dependent terms. Here we are only interested in $\alpha$ dependent terms. The decay distribution for this process contains a triple product correlation like the ones we have listed in Table~\ref{tab:obs} which one can get after summing over all the fermion spins in terms of $\epsilon_{\mu\nu\rho\sigma}p_{\tau^-}^\mu p_{\tau^+}^\nu p_{\pi^-}^\rho p_{\pi^+}^\sigma$. Here we have replaced neutrino momentum by $p_{\nu_\tau}=p_\tau - p_{\pi}$. 

We consider two different frames to study these momentum correlations. Observables $\mathcal O_1$ have been defined in $\tau^+\tau^-$ zero momentum frame (ZMF) in which both $\tau^+$ and $\tau^-$ are back-to-back (also known as ``Higgs rest frame''). In the ZMF frame, due to the large difference in the Higgs boson mass and the $\tau$ lepton mass, the $\tau^\pm$ are highly boosted leading to highly collimated decay products along the direction of $\tau^\pm$ momentum. This brings in some difficulty to reconstruct momenta of each particle in the event and hinders the prospects of performing angular analysis in such a frame. To get around these setbacks, we also define a peculiar frame where one part of the scalar product, constructed using tau momenta or  tau decay product momenta, is in the ZMF frame while the second part is constructed in $\tau^\pm$ rest frames (denoted as `prime' frame). In addition, the polarization is a frame dependent quantity and it depends on the Lorentz boost of the observable. Although there is no unique way that decides the frame, we have constructed some of our observables in the prime frame in order to extract the polarization information maximally.
Observables $\mathcal O_{2,3,4}$ are defined in this frame and the superscript $h$ or $\tau$ in the expression is to mark the corresponding rest frame. Note that one of the observable $\mathcal O_{2}$ was first introduced in~\cite{Berge:2008wi}, where efficiency was studied along with effects of cuts and smearing. Our results are consistent with that study. In addition, we also present an asymmetry as a function of the CP phase. 

\renewcommand{\arraystretch}{1.25}
\begin{table}[t]
\begin{center}
\newcolumntype{C}[1]{>{\centering\let\newline\\\arraybackslash\hspace{0pt}}m{#1}}
\begin{tabular}{ |C{5.1cm}|C{1.5cm}|  C{1.0cm} |}
\hline
Observables		& Frame\\\hline
$\mathcal O_1=(\vec{p}_{\tau^{-}} - \vec{p}_{\tau^{+}}).(\vec{p}_{\pi^{-}} \times \vec{p}_{\pi^{+}})$			& ZMF	\\\hline
$\mathcal O_2=(\vec{p}_{\tau^{-}} - \vec{p}_{\tau^{+}})^{h}.(\vec{p}_{\pi^{-}} \times \vec{p}_{\pi^{+}})^{\tau}$	& Prime	\\\hline
$\mathcal O_3=(\vec{p}_{\pi^{-}} - \vec{p}_{\pi^{+}})^{\tau}.(\vec{p}_{\pi^{-}} \times \vec{p}_{\pi^{+}})^{h}$		& Prime	\\\hline
$\mathcal O_4=(\vec{p}_{\pi^{-}} - \vec{p}_{\pi^{+}})^{h}.(\vec{p}_{\pi^{-}} \times \vec{p}_{\pi^{+}})^{\tau}$		& Prime\\
\hline
\end{tabular}
\caption{ T odd observables constructed in the process $h\to \tau^+\tau^-\to \pi^+\pi^-\nu\bar\nu$ at the LHC. All the observables have the definite CP and T transformation properties. Observables $\mathcal O_{1}$ have been defined in the Higgs rest frame or ZMF frame, while $\mathcal O_{2-4}$ are defined in prime frame (defined in the text).\label{tab:obs}}
\end{center}
\end{table}

\section{Reconstruction of semi-invisible event}
\label{sec:reconstruction}

In the previous section, we discussed the observables which are triple product correlations constructed out of momenta of $\tau^\pm$ and $\pi^\pm$ in several frames. However, to get any meaningful information on the usefulness of these CP observables at the LHC, it remains to be seen how precisely one can reconstruct these semi-invisible  tau pair events from the Higgs decay. Some of the recently proposed techniques in the literature particularly for such a scenario are in order. Popular and some of the early proposals, {\it viz.},   \textit{collinear approximation}~\cite{Ellis1988221,Maruyama:2015fis} determine invisible neutrino momenta by assuming tau decay products to be collinear. With this assumption, the neutrino(s) from tau take a fraction of tau momenta which results in the reduction of unknowns to two. Neutrino momenta then can be solved exactly using missing transverse momenta constraints. \textit{Missing mass calculator}~\cite{Elagin:2010aw,Xia:2016jec} solves for the four unknown components (explained in the next paragraph) of the neutrino momenta and remaining two unknowns are parametrized using a probability function. The probability function utilizes an independent measurement of angular separation between visible and invisible particles from $Z\rightarrow \tau \tau$ channel. \textit{Displaced vertex method}~\cite{Gripaios:2012th} assumes at least one tau decays via $3-$prong channel. It determines the tau momenta by utilizing the secondary vertex information and available constraints in the event. \textit{Constrained $\hat{s}$ method}~\cite{Swain:2014dha} assigns momenta to tau after optimizing the phase space by taking care of available kinematic constraints. \textit{$M_{2Cons}$ method}~\cite{Konar:2015hea,Konar:2016wbh} is a 3D $M_2$ variable which minimizes phase space by utilizing the Higgs mass and transverse momenta constraints and gives generic mass measurement prescription for antler decay topology. The reconstruction of neutrino momenta, in the present scenario, proved to be very precise. Recently developed reconstruction~\cite{Hagiwara:2016zqz} utilizes the tau mass-shell, missing transverse constraints together with measured impact parameter to reconstruct the semi-invisible events. The impact parameter is the perpendicular distance of pion momentum direction from the Higgs boson production vertex which can be identified using the tracks of jets produced with Higgs.

While there are many reconstruction methods, most of them are not sensitive to the observables considered in this analysis. A method which approximates the neutrino momenta exactly along the tau direction may not be sensitive to these variables because each of them is scalar triple products. In general, full reconstruction of these events is challenging because even for the hadronic decay of tau, there are two neutrinos present in the final state which traverse the detector without getting detected.
Full reconstruction of $h\to \tau^+\tau^-\to \pi^+\pi^-\nu_\tau \bar\nu_\tau$ requires determining all the components of the neutrinos momenta involved in the process. Assuming that there is no other source of missing energy in the process, the measured missing transverse momentum can be parametrized in terms of the unknown neutrinos momenta as follows,

\begin{eqnarray}
\slashed{E}_{Tx} &=& q_{\nu 1} \sin\theta_{\nu 1} \cos\Phi_{\nu 1} + q_{\nu 2} \sin\theta_{\nu 2} \cos\Phi_{\nu 2} \label{MET1},\\
\slashed{E}_{Ty} &=& q_{\nu 1} \sin\theta_{\nu 1} \sin\Phi_{\nu 1} + q_{\nu 2} \sin\theta_{\nu 2} \sin\Phi_{\nu 2}\label{MET2},
\end{eqnarray}
where $q_{\nu 1,2}$ are the magnitude of the neutrino momenta and $\theta_{\nu 1,\nu 2}$, $\Phi_{\nu 1,\nu 2}$ are polar and azimuthal angles of the neutrinos respectively. As it is evident from Eq.~\ref{MET1} and~\ref{MET2} that each tau pair event is under constraint because there are six unknowns but only five measurements including the tau and Higgs mass-shell constraints.
%
We start with two unknown degrees of freedom, $\Phi_{\nu 1}$ and $\Phi_{\nu 2}$, and scan over the full parameter space to completely solve the system with a mass window of 2 GeV for higgs boson which we attributed to the fact there can be, unknown measurement, error in it.
This resulted in fixing the angular separation between the corresponding visible and invisible particle but one parameter is still not constrained allowing the infinite number of solutions possible in each event.

At this point, it is important to note that the tau pair produced from Higgs decay possess considerably large decay length, as large as 87 $\mu m$, which we have not used so far. The large decay length of tau, in turn, allows for the measurement of the impact parameter which can then be used to get the tau direction. This additional measurement along with the other constraints discussed above uniquely determines the neutrinos momenta in each event\footnote{Note that the reconstruction proposed in ref.~\cite{Hagiwara:2016zqz} is slightly different than our method because it uses both the impact parameter of tau as well as the Higgs mass constraint to fully reconstruct the event. Since there are large measurement errors associated with the impact parameter, we took the one with larger magnitude for our analysis.}.   
%
%
To calculate the tau decay length vector first we take exact primary vertex coordinate from the {\tt Pythia} (v8)~\cite{Sjostrand:2006za} and calculated the impact parameter. We then smeared the primary vertex~\cite{Aaboud:2016rmg} using Gaussian smearing distribution with $\sigma_T = 0.01~ mm$ and $\sigma_Z = 0.1~mm$, in order to take the measurement error into account.
%
We pick that unique solution of $\Phi_{\nu 1}$ and $\Phi_{\nu 2}$ which gives the minimum error associated with the reconstructed tau decay length vector.

\section{Results}
\label{sec:sensitivity}

\begin{figure}[!h]\centering
\includegraphics[scale=0.585,angle=0,keepaspectratio=true]{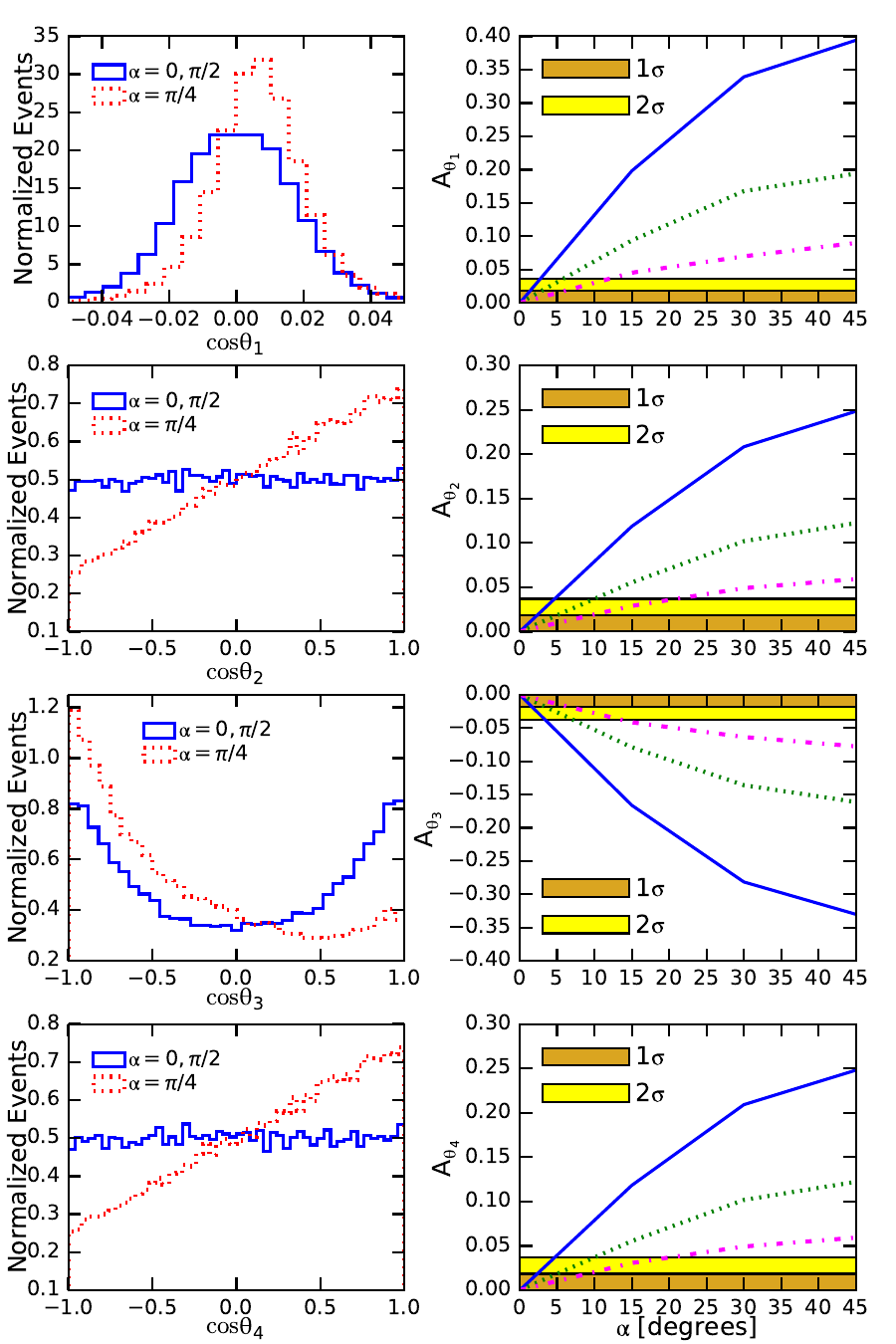}
\caption{Distribution of $\cos{\theta_i}$ for observables $\mathcal O_{1,\ldots,4}$ considering various values of CP violating phase $\alpha = 0, \pi/2$ for pure CP conserving and $\pi/4$  for a maximally CP violating Higgs at the LHC. Variations of corresponding asymmetries versus the phase $\alpha$ are presented in the right plot. The $1\sigma$ and $2\sigma$ bands for the statistical uncertainties (obtained using Eq. \ref{sd}) in the measurement of asymmetries with 1000 ${fb}^{-1}$ of integrated luminosity at the LHC are also shown. Solid (blue), dotted (green) and dash-dotted (magenta) curves denote the asymmetries obtained using the information at truth level, reconstruction level and reconstruction level with smearing of primary vertex respectively.}
\label{fig:dist_obs_cth}
\end{figure}

In this section, we discuss the statistical sensitivity of the observables proposed in this paper on the measurement of the CP phase of the tau Yukawa coupling. For the analysis, corresponding to each of the T odd observables $\mathcal O_{1,2,3,4}$, we focus on the angular correlations among the triple products listed in table \ref{tab:obs}, i.e., $\cos\theta_i=\hat{P}\cdot \hat{Q}$ where $P$ and $Q$ are first and second terms of the scalar-triple products.
 We display the distributions of  angular correlations $\cos\theta_{1,2,3,4}$ in Fig.~\ref{fig:dist_obs_cth} for a pure scalar ($\alpha=0$), pseudoscalar Higgs boson ($\alpha=\pi/2$) and for the maximally CP violating case ($\alpha=\pi/4$). As expected we find that the distribution is symmetric for $\alpha=0$ and ($\alpha=\pi/2$) which denote the CP conserving scenarios and thus leading to the vanishing expectation values of the asymmetries (defined in Eq.~\ref{Asymmetry}) associated with these observables. On the other hand, for a maximally CP violating scenario i.e., $\alpha=\pi/4$, there is a significant distortion in the distributions relative to CP conserving case indicating that the asymmetries are sensitive to CP violating phase $\alpha$.

For each distribution shown in the Fig.~\ref{fig:dist_obs_cth}, we define a corresponding asymmetry as follows
\beq\label{Asymmetry}
A_{\theta_i}=\frac{1}{\mathcal N_{\rm tot}} \left[\mathcal N(\cos\theta_i<0)-\mathcal N(\cos\theta_i>0)\right]
\eeq
where $\mathcal N_{\rm tot}$ is the total number of events. Note that the expression for the asymmetry can be applied for both the new physics contributions and the SM backgrounds. Including the contributions from the SM background, the total asymmetry, $A_{\theta_i}^{\rm total}$, can be written as

\begin{eqnarray}\label{Asym_tot}
 A_{\theta_i}^{\rm total}&=&A_{\theta_i}^{\rm NP}R + A_{\theta_i}^{\rm bckg}(1-R),\\
 R&=&\frac{\sigma_{\rm NP}}{\sigma_{\rm NP} + \sigma_{\rm bckg}},
\end{eqnarray}
where $\sigma_{\rm NP}$, $\sigma_{\rm bckg}$, $A_{\theta_i}^{\rm NP}$ and $A_{\theta_i}^{\rm bckg}$ are the contributions to new physics (NP) cross section, background cross section, asymmetry due to NP and asymmetry due to SM backgrounds respectively. From the Eqns. \ref{Asym_tot}, we can see that the effect of the background contribution to the asymmetry is to reduce the magnitude of $R$ and thus reducing the sensitivity of NP contribution to the asymmetry.

We also study the behavior of these asymmetries as a function of CP phase $\alpha$. These asymmetries have been displayed in the right panels of Fig.~\ref{fig:dist_obs_cth} for observables $\mathcal O_{1,2,3,4}$ respectively. The blue (solid) curve for the asymmetry in the figure denote the truth-level scenario assuming that information regarding the tau momenta is fully known. The green (dash-dotted) curve denote the case where the tau momenta have been reconstructed using the formalism discussed in section \ref{sec:reconstruction}. The magenta (dotted) curve is obtained when the reconstruction of tau moments is performed along with the smeared primary vertex and thus presents the most realistic estimation of the asymmetry in an actual LHC environment.

From the plots of asymmetries, we find that the asymmetry is vanishing for a CP conserving scenario ($\alpha=0$ or $\pi/2$) resulting from a symmetric $\cos\theta_i$ distribution. For a maximally CP violating scenario ($\alpha=\pi/4$), the asymmetry is the largest for the observable $\mathcal O_3$ with the value $\sim 33\%$ (at truth-level) while observables $\mathcal O_{2,4}$ also provide the modest asymmetry of $\sim 25\%$. Also, the slopes of the asymmetries are fairly steep showing a good sensitivity to the measurement of CP phase $\alpha$. However, at the realistic scenario (after the reconstruction of taus momenta and smeared primary vertex), the asymmetry drops somewhat as can be seen from the figure. Nevertheless, asymmetries are significant enough to provide stringent bounds on the CP angle $\alpha$. 

Looking at $\mathcal O_{2}$ and $\mathcal O_{4}$, one can realize that both are essentially the similar variables with first term  replacing tau momenta with the corresponding pion momenta at the Higgs rest frame. In general, they can generate different contributions. However, in our present example, tau's are highly boosted at the Higgs rest frame. Thus, the decaying pion would essentially follow almost the same direction as corresponding tau and hence generating nearly close values both in these variables\footnote{Here we emphasize that these observables can also be interesting, depending on the boost of the daughter particle, in other scenarios as well. For example the heavy Higgs, in a BSM scenario, to top pair process can be a potential channel to apply these variables where one does not expect the top quark to be highly boosted unlike the tau case, as a result the observable $\mathcal O_2$ and $\mathcal O_4$ would behaves differently.}. This is also evident in the angular distribution and asymmetry.

We now discuss the sensitivity of these asymmetries to the measurement of CP phase, $\alpha$, in $H\tau^+\tau^-$ coupling at the 13 TeV LHC. To obtain the bound on CP violating coupling $\alpha$, we find those values of $\alpha$ for which the asymmetries deviate from the SM prediction by a certain confidence level. The statistical uncertainty in the measurement of an asymmetry is defined as follows
\begin{equation}\label{sd}
 \Delta \mathcal A=\frac{\sqrt{1-\mathcal A_{\rm SM}^2}}{\sqrt{\sigma_{\rm SM} ~\epsilon~ {\mathcal L} }},
\end{equation}
where  $\mathcal L$ is the integrated luminosity, $\mathcal A_{\rm SM}$ is the expected value of an asymmetry in the SM, $\sigma_{\rm SM}$ is the total tau pair cross section in the SM and $\epsilon$ is the experimental efficiency factor for the detection of such events after inclusion of realistic cuts and background elimination. We estimate this efficiency utilizing the recent analysis on Higgs boson searches in its hadronic $\tau$ decays. The $\epsilon$ is the ratio between the number of events after the realistic cuts and the expected number of events. From recent ATLAS paper ~\cite{Aad:2015vsa} for gluon fusion channel and subsequent decays of Higgs into hadronic taus, the efficiency factor turned out to be 8.9\%. The expected number of events is obtained by the product of the theoretical Higgs production cross section in gluon fusion at 8 TeV (19.27 pb), the Higgs decay branching ratio into tau pairs, tau decay branching fractions to charged pion and the integrated luminosity. We have assumed this efficiency factor to be same for 13 TeV and used in the estimation of sensitivity of our observables.

In the right panel of Fig.~\ref{fig:dist_obs_cth}, we display the $1 \sigma$ and  $2 \sigma$ statistical uncertainty in the measurement of respective asymmetries through shaded bands. While presenting these statistical regions, we consider 1000 ${fb}^{-1}$ of integrated luminosity. From the figures, we find that the asymmetry $A_{\theta_1}$ is the most sensitive of all the asymmetries we analyzed in this analysis and the measurement of this asymmetry can determine the CP phase $\alpha$ up to 12 degrees at 2$\sigma$ CL. The asymmetries $A_{\theta_2}$, $A_{\theta_3}$ and $A_{\theta_4}$ can determine this angle up to 20, 15 and 20 degrees, respectively at 2$\sigma$ CL for 13 TeV LHC.

\begin{figure}[!h]\centering
\includegraphics[scale=0.24,angle=0,keepaspectratio=true]{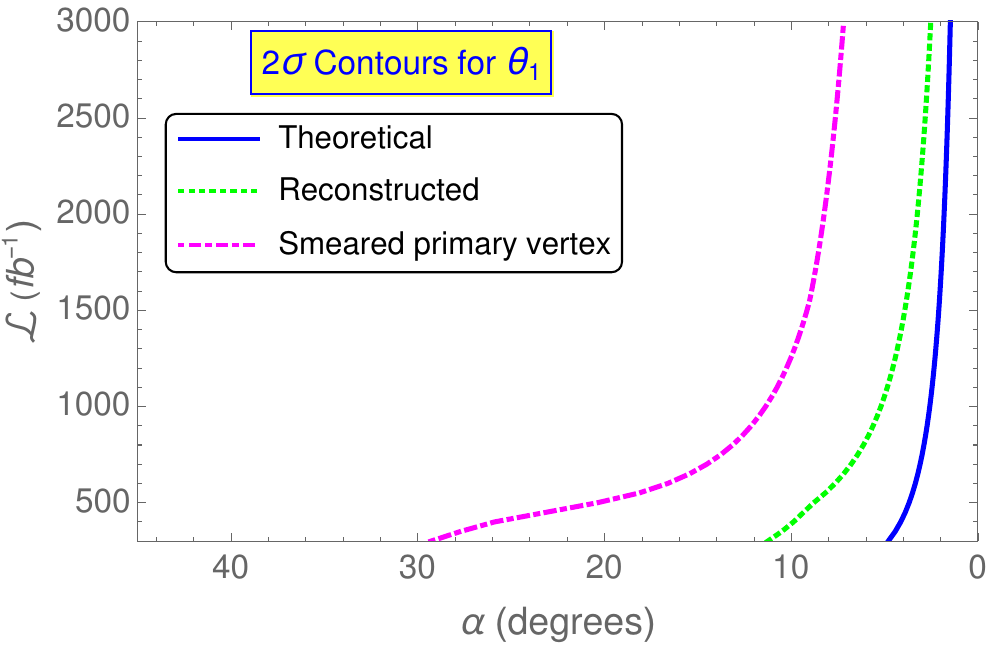}
\includegraphics[scale=0.24,angle=0,keepaspectratio=true]{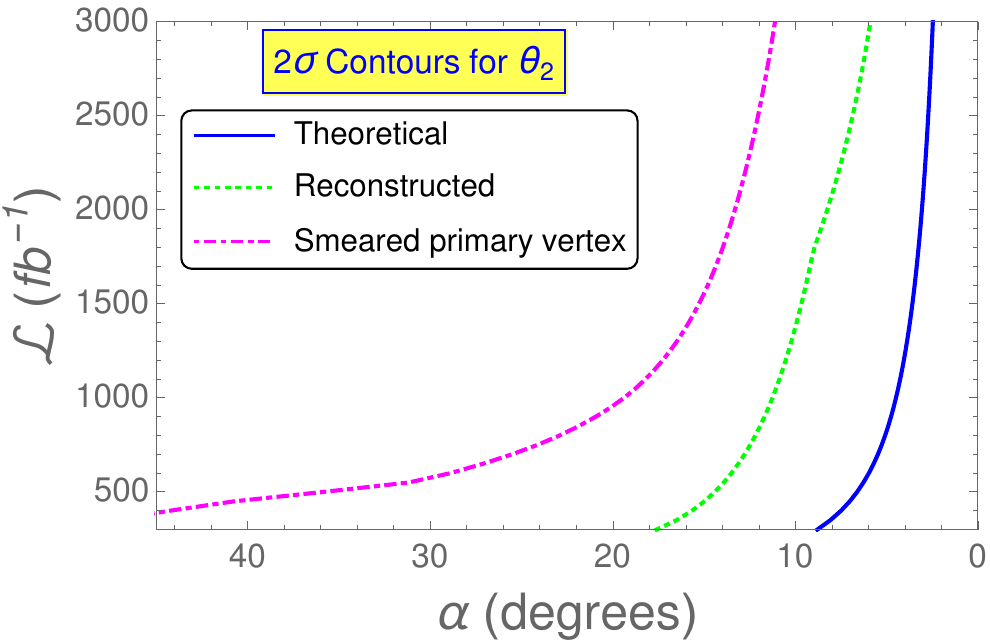}
\includegraphics[scale=0.24,angle=0,keepaspectratio=true]{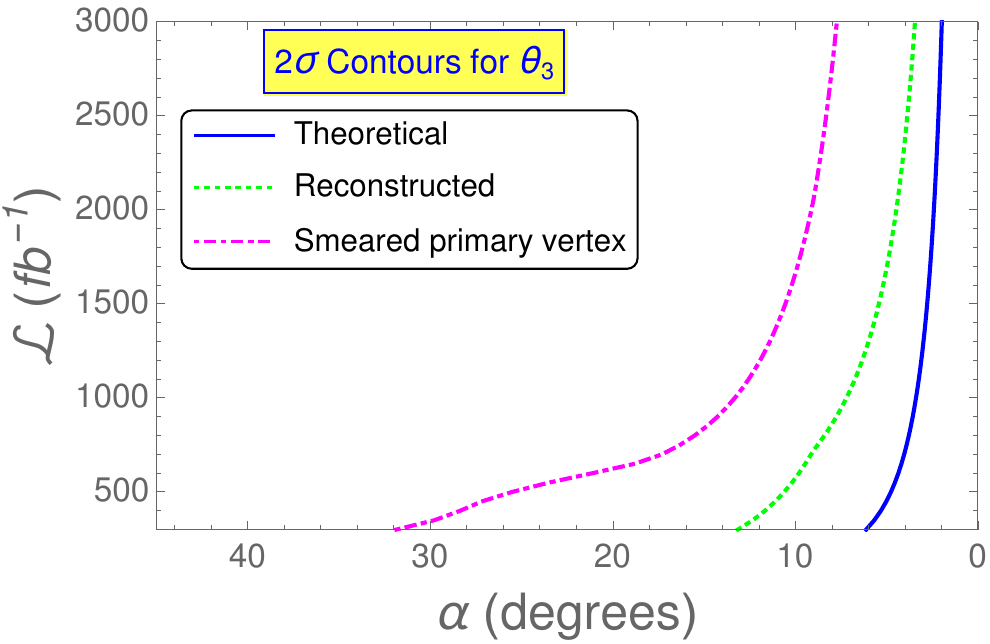}
\includegraphics[scale=0.24,angle=0,keepaspectratio=true]{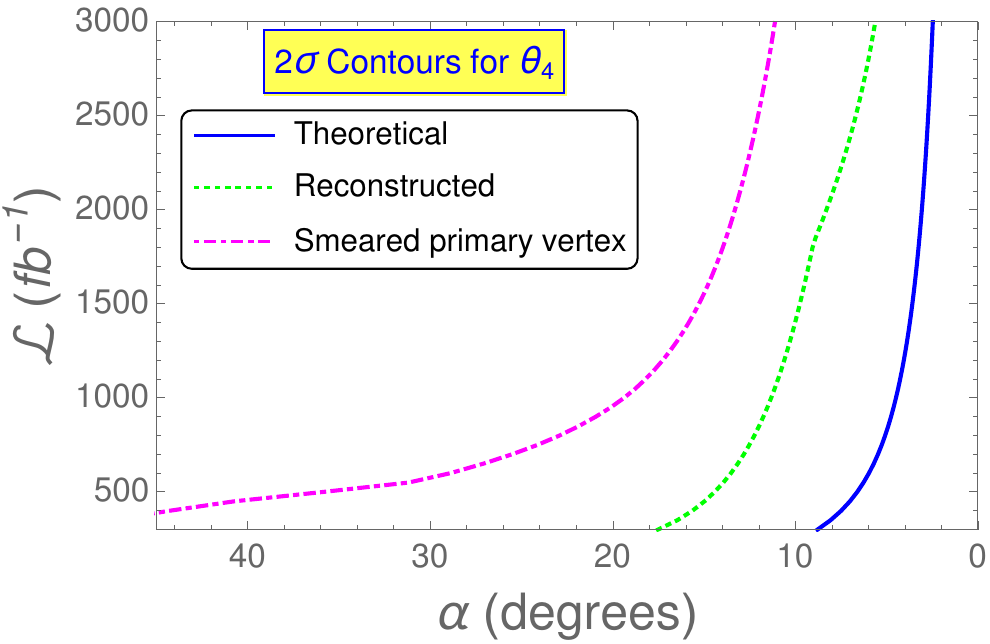}
\caption{The figure shows 2$\sigma$ statistical sensitivity for the CP phase $\alpha$ which can be pinned down with the increasing luminosity at the HL-LHC. The 2$\sigma$ statistical sensitivity is calculated using information at truth level (blue solid), reconstruction level (green dotted) and reconstruction level with the smearing of primary vertex (magenta dash-dotted) respectively.}
\label{fig:lum}
\end{figure}

In the Fig.~\ref{fig:lum} we present the 2$\sigma$ statistical sensitivities of the observables up to which the CP phase $\alpha$ can be probed with the projected luminosity for 14 TeV at the HL-LHC. The sensitivity of each observable at the theoretical level (blue solid), reconstructed level (green dotted) and the reconstructed with the smeared primary vertex (magenta dashed-dotted) are displayed as a function of integrated luminosity ranging from 300 to 3000 ${fb}^{-1}$. We find that the asymmetry $A_{\theta_1}$ can pin down the CP phase $\alpha$ up to 6 degree for 3000 ${fb}^{-1}$ at 14 TeV HL-LHC. Here we have taken the tau momentum reconstruction efficiency into account along with uncertainty in the primary vertex as well as the efficiency factor $\epsilon$ for this channel.

Note that while estimating the bounds on the CP violating phase $\alpha$ using various asymmetries, we do not include background contribution into the total asymmetry as defined in Eqns. \ref{Asym_tot}. Thus, in this analysis, we assume $R$ to be equal to 1. Thus, it is obvious that in the presence of background, the sensitivity of the various observables would reduce. Nevertheless, the focus of this work is to suggest some new observables for extracting CP phase in tau-Yukawa couplings and present a new reconstruction technique for tau-momentum at the LHC. 

\section{Summary and Conclusion}
\label{sec:conclusion}

The determination of the CP properties of Higgs boson is one of the important aims at the large hadron collider (LHC) in its current and future runs. The goal is facilitated in the Higgs couplings to the third generation of fermions, in particular $\tau^\pm$ leptons. Spin of $\tau^\pm$ and the correlations between them may provide a great insight to the CP properties of Higgs boson. However, these are not directly measurable and manifest themselves in the distribution of its decay products.

In spirit of the aforementioned fact that the spin correlations are reflected in final state distributions, we proposed several triple product correlations which are constructed from the momenta of various particles involved in the process. Recognizing that the sensitive observables are best represented at the rest frame, we consider two different type of frames to study the correlations. These correlations have a definite CP and T transformation properties. We present the distribution of angular correlations obtained the various momentum correlations discussed earlier. These are shown to be sensitive to the CP phase in the $H\tau^+\tau^-$ couplings at the LHC. 

In section \ref{sec:reconstruction}, we discussed various methods of tau momentum reconstruction available in the literature. We also proposed a new method of tau reconstruction which is based on the measurements of impact parameter and primary vertex. We employed this method to reconstruct the various observables studied in the paper and analyzed the distribution and asymmetries in the realistic LHC environment.

We also constructed the asymmetries using each angular correlation and studied their behavior as a function of the CP phase. Some of these asymmetries are found to be as large as 35\% for the maximally CP violating scenario. A statistical analysis of the sensitivity of these asymmetries on the measurement of CP phase is studied with the reconstruction efficiency of $\tau^\pm$ pair events at the LHC. We found that with 1000 ${fb}^{-1}$  of integrated luminosity, the CP phase can be determined up to 15 degrees at the 13 TeV LHC.

%


\bigskip
\section*{Acknowledgments}
This work is supported by Physical Research Laboratory (PRL), Department of Space, Government of India, with computational support from Vikram-100 HPC at PRL. P.S. acknowledges the support from the University of Adelaide and the Australian Research Council through the ARC Center of Excellence for Particle Physics (CoEPP) at the Terascale (grant no.\ CE110001004).


\bibliographystyle{unsrt}
\bibliography{bibliography}

\end{document}